\documentclass{article}

\usepackage{arxiv}

\usepackage[utf8]{inputenc} 
\usepackage[T1]{fontenc}    
\usepackage{hyperref}       
\usepackage{url}            
\usepackage{booktabs}       
\usepackage{amsfonts}       
\usepackage{nicefrac}       
\usepackage{microtype}      
\usepackage{lipsum}
\usepackage{algorithm}
\usepackage{algorithmic}
\usepackage{graphicx}

\title{J-Parallelio - automatic parallelization framework for java virtual machine code}

\author{
  Krzysztof Stuglik \\
  Institute of Computer Science\\
  University of Science and Technology\\
  Cracow, Poland \\
  \texttt{stuglik@student.agh.edu.pl} \\
   \And
 Piotr Listkiewicz \\
  Institute of Computer Science\\
  University of Science and Technology\\
  Cracow, Poland \\
  \texttt{listkiewicz@student.agh.edu.pl} \\
   \And
 Mateusz Kulczyk \\
  Institute of Computer Science\\
  University of Science and Technology\\
  Cracow, Poland \\
  \texttt{kulczyk@student.agh.edu.pl} \\
   \And
 Marcin Pietron \\
  Institute of Electronics\\
  University of Science and Technology\\
  Cracow, Poland \\
  deep-cogni.com \\
  \texttt{pietron@agh.edu.pl} \\
}

\begin{document}
\maketitle

\begin{abstract}
Manual translation of the algorithms from sequential version to its parallel counterpart is time consuming and can be done only with the specific knowledge of hardware accelerator architecture, parallel programming or programming environment. The automation of this process makes porting the code much easier and faster. The key aspect in this case is how efficient the generated parallel code will be. The paper describes J-Parallelio, the framework for automatic analysis of the bytecode source codes and its parallelisation on multicore processors. The process consists of a few steps. First step is a process of decompilation of JVM and its translation to internal abstract syntax tree, the dependency extraction and memory analysis is performed. Finally, the mapping process is performed which consists of a set of rules responsible for translating the input virtual machine source code to its parallel version. 
The main novelty is that it can deal with pure java virtual machine and can generate parallel code for multicore processors. This makes the system portable and it can work with different languages based on JVM after some small modifications. The efficiency of automatically translated source codes were compared with their manually written counterparts on chosen benchmarks.
\end{abstract}


\section{Introduction}
Over the last few years we have observed a lot of trials of building tools that can help with automatic code parallelisation on different hardware platforms. We have also observed intensive research in parallelisation of source codes written in C/C++ language and speed up using OpenMP and OpenCL environments \cite{IEEEhowto:bondhugula}, \cite{IEEEhowto:amini}, \cite{IEEEhowto:liao}. Recently, we have been able notice that tools for transforming C to CUDA or OpenCL for GPU acceleration are a common topic of interest in automatic code parallelisation \cite{IEEEhowto:baskaran}, \cite{IEEEhowto:elmqvist}, \cite{IEEEhowto:setoain}, \cite{IEEEhowto:amini} and \cite{IEEEhowto:hou}. There were also trials to automatically speed up different languages (e.g. Java \cite{IEEEhowto:rafael}) on multicore processors.
In the J-Paralellio system, a fairly new approach is presented. The framework is fully based on java virtual machine code and language independent, fig. \ref{fig:parallelio}. The main input is java virtual machine code. JVM is parsed and partially decompiled and transformed to intermediate structures, then analysed, instrumented and transformed back to virtual machine in a  parallelised version. 
The crucial element of its functionality is an engine incorporated with a decompiler and abstract syntax tree builder, pool of algorithms responsible for dependency and memory access analysis and set of rules for java virtual machine code transformation. 
In the first stage, the system parses JVM source code and extracts loops from the JVM code. It then generates an Abstract Syntax Tree with Data Flow Graph which helps to identify the potential parallelism by extracting internal dependencies. The system mainly concentrates on loops analysis. 
The framework performs the dependency analysis using the same algorithms as other automatic paralellisation systems \cite{IEEEhowto:bondhugula}, \cite{IEEEhowto:baskaran}, \cite{IEEEhowto:ppcg}, \cite{IEEEhowto:amini}.
Finally, the system maps the loops iterations and data structures to the hardware accelerator by using built-in translation rules. After the system engine analysis, it generates a parallel version of java virtual machine code. It instruments java virtual machine code with instructions responsible for the creation of threads. It automatically maps the loops to multi-thread execution. 
Finally, the work presents the results of mapping a few benchmarks from sequential to parallel versions.
The main advantage of the presented approach is its portability. There are several languages which are based on Java virtual machine. JRuby and Jython are perhaps the most well-known ports of existing languages (Ruby and Python respectively). The new languages that have been created from scratch to compile to Java bytecode, Clojure, Groovy and Scala may be the most popular examples.

The paper is organised as follows: the second section describes the related works in the field of automatic parallelisation. The next sections concentrates on methodology and algorithms which analyse and translate the parallel version of the input sequential JVM source code. The fifth section presents the results, the sixth conclusions and future work. 

\section{Related works}
The articles \cite{IEEEhowto:rafael}, \cite{IEEEhowto:sun} and \cite{IEEEhowto:bradel} present approaches for automatic java source code parallelisation. In \cite{IEEEhowto:rafael} dependency extraction methods are used for java source code analysis. The translator is built for sequential JAVA code which generates a highly parallel version of the same program. The translation process interprets the AST nodes for signatures such as read-write access, execution-flow modifications, among others and generates a set of dependencies between executable tasks. The presented approach has been applied for recursive Fibonacci and FFT algorithms. The methods obtained a 10.97x and 9.0x increase in speed on a twelve-core machine.
The latter two methods \cite{IEEEhowto:sun} and \cite{IEEEhowto:bradel} concentrate on parallelisation using trace information. The approach presented in \cite{IEEEhowto:sun} collects on-line trace information during program execution, and dynamically recompiles methods that can be executed in parallel. In \cite{IEEEhowto:bradel}, the authors implement a system that demonstrates the benefits and addresses the challenges of using traces for data-parallel applications. They propose an execution model for automatic parallelisation based on traces. In \cite{IEEEhowto:chan}, a novel approach is described and evaluated for the automatic parallelisation of programs that use pointer-based dynamic data structures written in Java. The approach exploits parallelism among methods by creating an asynchronous thread of execution for each method invocation in a program. The only work in which java code parallelisation is done directly on a java virtual machine is shown in \cite{IEEEhowto:felber}. It is semi-automatic, there is no detailed JVM code analysis, it has not decompilation, automatic JVM code transformation and generation. It presents the process of the automatic instrumentation of virtual machine code by preparing and invoking special adapters which can run the original methods in a multi-threaded java environment. The results are presented using the Mandelbrot benchmark. The drawback of the article is the lack of results of running the algorithm on more benchmarks.  

Several other systems were designed for automatic code parallelisation which are mainly based on C/C++ language. YUCCA \cite{IEEEhowto:smitha} designed by KPIT Technologies is an automatic parallelisation tool for projects written in C language. It provides source to source conversion - on input, it takes the source code of the application written in C and produces a parallelised version of the source code as an output. YUCCA output is a multithreaded version of the input with Pthreads or OpenMP pragmas inserted at appropriate places. YUCCA uses PThreads to perform task parallelisation and OpenMP to make loops run in parallel. YUCCA consists of two main parts: the front-end, which is responsible for parsing source code, the back-end which performs static dependency analysis to identify parts of code that is worth being parallelised. 
PLUTO \cite{IEEEhowto:bondhugula}, \cite{IEEEhowto:bondhugula2008} is an automatic parallelisation tool based on a polyhedral model. PLUTO performs source to source transformation - it conducts coarse-grained parallelism and at the same time ensures data locality. The core transformation framework mainly works by finding affine transformations for efficient tiling. PLUTO performs parallelisation with OpenMP and the code is also transformed for locality. The tool provides a number of options to tune aspects such as tile sizes, unroll factors and outer loop fusion structure.
C-to-CUDA \cite{IEEEhowto:baskaran} and PPCG \cite{IEEEhowto:ppcg} propose similar steps to solve the automatic GPGPU code-generation problem. They concentrate on finding parallel loops, the creation of a polyhedral model from the loops; they tile and map the loops to GPU blocks and threads and determine where to place the data.

Par4All \cite{IEEEhowto:amini} is an automatic parallelising and optimising compiler for C and Fortran sequential programs. The purpose of this source-to-source compiler is to adapt existing applications to various hardware targets such as multicore systems, high performance computers and GPUs. It creates a new source code and thus allows the original source code of the application to remain unchanged.
The auto-parallelisation feature of the Intel C++ Compiler \cite{IEEEhowto:intel_c} automatically translates serial portions of the input program into semantically equivalent multi-threaded code. Automatic parallelisation determines the loops that are good candidates, performs the data-flow analysis to verify correct parallel execution, and partitions the data for threaded code generation as is needed in programming with OpenMP directives. The OpenMP and auto-parallelisation applications provide the performance gains from shared memory on multiprocessor systems. 
AutoPar \cite{IEEEhowto:liao} is a tool which can automatically insert OpenMP pragmas into input serial C/C++ codes. For input programs with existing OpenMP directives, the tool double checks the correctness when the right option is turned on. Compared to conventional tools, AutoPar can incorporate user knowledge (semantics) to discover more parallelisation opportunities.

The iPat/OMP \cite{IEEEhowto:ishikara} tool provides users with the assistance needed for the OpenMP parallelisation of a sequential program. This tool is implemented as a set of functions on the Emacs editor. All the activities related to program parallelisation, such as selecting a target portion of the program, invoking an assistance command, and modifying the program based on the assistance information shown by the tool, can be handled in the source program editor environment. OMP2MPI \cite{IEEEhowto:garriga_} automatically generates MPI source code from OpenMP, allowing the program to exploit non shared-memory architectures such as cluster, or Network-on-Chip-based (NoC-based) Multiprocessors-System-on-Chip (MPSoC). OMP2MPI provides a solution that allows further optimisation by an expert who wants to achieve better results. 


\section{Methodology}
The framework consists of a few submodules. The first is a decompilation comprised of AST building components. It is responsible for transforming the Java bytecode to Java instructions and building an Abstract Syntax Tree from them. In parallel with decompilation, the loops extraction module works. It enables extraction of the loops in a java bytecode. The loops are the analysed by a specialised algorithm to extract potential parallelism (see Section 3.1).  
After analysis, mapping the bytecode to a multithread version is performed. The bytecode is instrumented with special instructions which are responsible for the thread, the task and their memory management. The multithreaded bytecode can then be run or decompiled to any language based on a java virtual machine.  
The main phases of the framework method are: 

\begin{figure*}
\centering
\includegraphics[scale = 0.8]{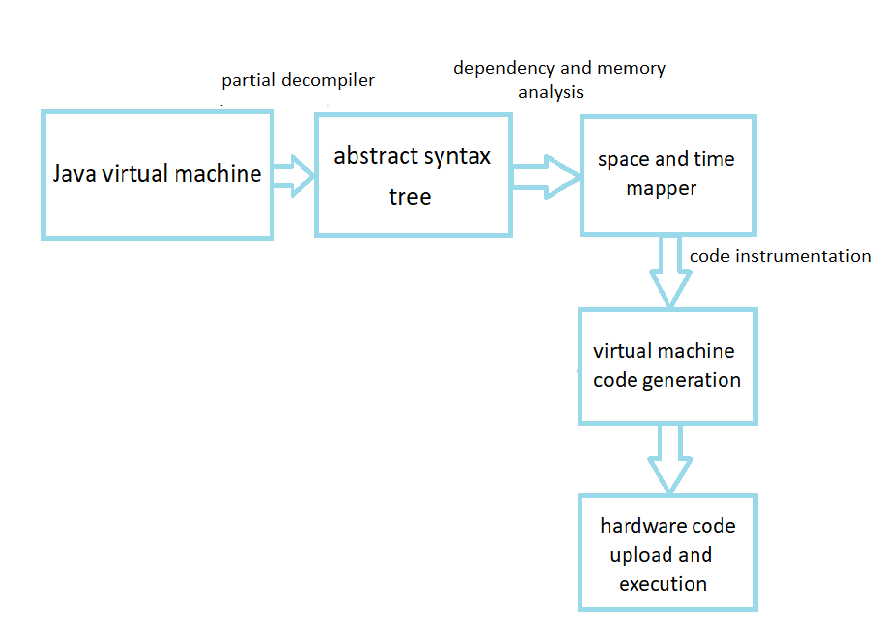}
\caption{\footnotesize Automatic java virtual machine parallelization system.}
\label{fig:parallelio}
\end{figure*}

\begin{itemize}
\item Decompilation
\item Abstract syntax tree building
\item Loops extraction
\item Data flow and dependency analysis  
\item Multithread mapping and JVM code instrumentation
\end{itemize}

\subsection{Java virtual machine and BCEL}
A Java virtual machine (JVM) is an abstract computing machine that enables a computer to run a Java program or program in other language which is based on JVM (e.g. Clojure, Groovy or Scala). One of the organisational units of JVM bytecode is a class. The JVM has instructions for the following groups of tasks: arithmetic operations, load and store arithmetic, type conversion, object creation and manipulation, stack management (stack operations push / pop), control transfer (branching), field access, method invocation, throwing exceptions and monitor-based concurrency. The JVM operation set can be represented as: $OP\ ARG$, where $OP$ belongs to set of JVM available elementary operations, $ARG$ is an argument. Argument can be constant or variable. In each single instruction can be one or two arguments. The bytecode instruction set currently consists of 212 instructions and 44 opcodes. 

The presented framework uses ByteCode Engineering Library (BCEL) which enables reading and manipulating in Java bytecode. The BCEL is intended to give users a convenient way to analyse, create, and manipulate Java class files. Classes are represented by objects which contain all the symbolic information of the given class: methods, fields and bytecode instructions, in particular.
Such objects can be read from an existing file, transformed by a program (e.g. a class loader at run-time) and written to a file again. 




\subsection{Decompilation and AST building}

The first stage of the framework is responsible for translation of the raw Java bytecode to higher level instructions. Next, the translated instructions should be analysed to extract dependency between program components. Therefore the algorithm was built to fulfill this goal. 
Each single bytecode operation can pop or push elements on the JVM stack. Each single instruction in high-level language ends when the stack is empty. Therefore to extract whole Java instruction the module monitors the state of the stack.
The approach is presented in Algorithm \ref{alg:decomp}. It loads a class implementation in Java bytecode and gets a list of its methods (line 1). It also initializes the stack of the virtual machine ($S$) and decompiled instruction collections ($J_{i}$ and $J$).
 
\begin{algorithm}[H]
\begin{algorithmic}[1]
\STATE {$S$ $\gets$ $\emptyset$, $J_{i}$ $\gets$ $\emptyset$, $J$ $\gets$ $\emptyset$, $C$ $\gets$ load class, $M_{C}$ $\gets$ get methods in $C$} 
\FOR {$m$ \textbf{in} $M_{C}$}
\STATE {$I_{set}$ $\gets$ get instructions from $m$}
\WHILE {$I_{set}$ $\notin$ $\emptyset$}
\STATE {$i$ $\gets$ remove first instruction from $I_{set}$}
\STATE {$push$ $i$ to $S$}
\WHILE {$S$ $\notin$ $\emptyset$}
\STATE {$i$ $\gets$ remove first instruction from $I_{set}$} 
\IF {$i$ $\in$ $OP$}
\STATE {$ARG$ $\gets$ $pop$ $ARG$ from $S$}
\STATE {$R$ $\gets$ $OP$ $ARG$}
\STATE {$push$ $R$ to $S$}
\ENDIF
\IF {$i$ $\in$ $PUSH$}
\STATE {$push$ $ARG$ to $S$} 
\ENDIF 
\IF {$i$ $\in$ $POP$}
\STATE {$J_{i}$ $\gets$ $pop$ from $S$}
\ENDIF
\ENDWHILE
\STATE {return $J_{i}$}
\STATE {$J$ $\gets$ $J_{i}$ $\cup$ $J$}
\ENDWHILE
\ENDFOR
\end{algorithmic}
\caption{Decompilation the JVM}
\label{alg:decomp}
\end{algorithm}

Then the algorithm iterates over the class methods. It gets an instruction list from each method $m$ (line 3). It translates each instruction in a sequence and monitors the state of the stack. At the beginning it pushes the first instruction $i$ of $J_{i}$ to $S$ (line 6). In a while loop it processes next bytecode instructions until stack $S$ is empty. If it recognizes the operand instruction then it takes arguments of operation from the stack (line 10).
It forms the triple address instruction (line 11) and pushes it to the stack. If $PUSH$ operation is recognized (line 14) it pushes the argument of bytecode instruction to the stack. Finally if a single $POP$ instruction is met the algorithm takes the final decompiled instruction ($J_{i}$) from the stack (line 18). All decompiled instructions are stored in $J$ list (line 22). 

The Algorithm \ref{alg:ast} presents building Abstract Syntax Tree from the list of the decompiled instructions. It initializes data structures (line 1) in which it stores dependencies between instructions ($T$) and list of assigned variables ($L$). Then, it goes through the list of instructions and takes left and right hand variables from the analysed instruction (line 3-5). It checks in a loop if every right hand variable of $J_{i}$ (from $J_{iR}$) already exists in $L$. If it is true, the algorithm takes the last instruction in which the specified right hand variable has appeared ($var$, line 8).

\begin{algorithm}
\begin{algorithmic}[1]
\STATE{$T$ $\gets$ $\emptyset$, $L$ $\gets$ $\emptyset$}
\WHILE{$J$ $\notin$ $\emptyset$}
\STATE{$J_{i}$ $\gets$ remove first instruction from $J$}
\STATE{$J_{iR}$ $\gets$ get right hand variables of $J_{i}$}
\STATE{$J_{iL}$ $\gets$ get left hand variable of $J_{i}$}
\FOR{$var$ \textbf{in} $J_{iR}$}
\IF{$var$ $\in$ $L$}
\STATE{$I_{R}$ $\gets$ take last instruction from $L$: $var$ $\in$ $I_{R}$}
\STATE{$d$ $\gets$ ($I_{R}$ $\rightarrow$ $J_{i}$)}
\STATE{$T$ $\gets$ $T$ $\cup$ $d$}
\ENDIF
\ENDFOR
\STATE{$L$ $\gets$ $J_{iR}$ $\cup$ $L$}
\ENDWHILE
\STATE{return T}
\caption{AST building}
\label{alg:ast}
\end{algorithmic}
\end{algorithm}

The dependency $d$ is extracted between these two instructions ($I_{R}$ and $J_{i}$, line 9) and added to the set $T$. After the instruction is processed the right hand variable is added to $L$ set (line 13). At the end of the algorithm set of dependencies $T$ is returned (line 15) which can be directly used to create the AST.  

\subsection{Loops extraction and data flow analysis}

The loop extraction and their analysis are the next step of Java bytecode analysis. It is a crucial stage in the automatic parallelisation because loops are the main source of hidden parallelism. The Algorithm \ref{alg:loop} presents how loops are extracted from the bytecode. The process starts from finding jump $GOTO$ instruction. Then the argument of jump instruction is read (jump address, line 2). The address is the start of the program loop. The algorithm goes to this location and parses the iteration variable with its initialization value and loop boundary condition. It takes the list of the instruction from the location just after conditional bytecode instruction to the $GOTO$ instruction (line 6). The extracted body of the loop can be decompiled using Algorithm \ref{alg:decomp}. 
Very often loops can be nested. Therefore to extract hierarchy of the loops presented method should run recursively. 

\begin{algorithm}
\begin{algorithmic}[1]
\STATE{$I_{goto}$ $\gets$ find $GOTO$ instruction}
\STATE{$L_{s}$ $\gets$ take address from $I_{goto}$ instruction}
\STATE{jump to $L_{s}$}
\STATE{$i$ $\gets$ take iteration variable}
\STATE{$B$ $\gets$ take condition boundary of the loop}
\STATE{$L$ $\gets$ take block from condition to $I_{goto}$}
\STATE{decompile loop $L$  //Algorithm \ref{alg:decomp}}
\caption{The JVM loop extraction}
\label{alg:loop}
\end{algorithmic}
\end{algorithm}

After loop extraction the loop analysis is performed. Formal analysis is based on a polyhedral model; algorithms for dependency detection are run by using symbolic Fourier-Motzkin elimination. 

\section{Java virtual machine automatic parallelisation module}


The Algorithm \ref{alg:jvm_p} describes the process of instrumenting the JVM bytecode. The first two steps are responsible for initialization the thread executors and tasks list (line 1 and 2). The next part depends on type of parallelism. If data-driven
dependency is recognized (lines 3-7, e.g. histogram) the input data is divided to independent data chunks (line 4). Output data is copied and create separate instance for each parallel thread (line 5). Then subtasks methods are created to work on these data chunks (line 6). At the end the single thread code responsible for merging result data is added (line 7). 

\begin{algorithm}
\begin{algorithmic}[1]
\STATE{$Ex$ $\gets$ create executors}
\STATE{[$T_{1}$, $T_{2}$,..., $T_{N}$] $\gets$ create task list}
\IF{$P_t$ \textbf{is} $DP$}
\STATE{[$D_{i1}$, $D_{i2}$, ..., $D_{iN}$] $\gets$ divide the input data $D_{i}$}
\STATE{[$D_{o1}$, $D_{o2}$, ..., $D_{oN}$] $\gets$ make copy of output data $D_{o}$}
\STATE{$P$ $\gets$ add pool of subtasks ($D_{in}$, $D_{on}$)}
\STATE{$D_{o}$ $\gets$ add merging output data $D$}
\ENDIF
\IF{$P_t$ \textbf{is} $IP$}
\STATE{[$(start_{1}$, $step_{1}$), ($start_{2}$, $step_{2}$), ..., ($start_{N}$, $step_{N}$)] $\gets$ divide iterations to chunks}
\STATE{$P$ $\gets$ add pool of subtasks in parallelised regions ($start_{n}$, $step_{n}$)}
\ENDIF

\caption{The JVM parallelisation}
\label{alg:jvm_p}
\end{algorithmic}
\end{algorithm}

The JVM instrumentation is finally used in the main parallelisation algorithm which is shown in Algorithm \ref{alg:par}. The algorithm tries to find best parallel configuration of the input sequential program. The parallelisation concentrates mainly on program loops extracted by Algorithm \ref{alg:loop} (line 2). Then the dependency analysis is performed (line 3). The loops can be fully or partially parallelised or unable to run in parallel (line 4). In the case of fully parallel loops the main decision is which loop to choose in a nested loop structure. 
Additionally, loops can be transformed by interchanging, tiling, skewing etc. By making appropriate selections from this choice of transformations it is possible to achieve better mapping and more efficient implementation. All these configurations are generated in line 5. Then they are tested in a loop (lines 6-10). Each candidate transformation is parallelised in JVM. Next, they are run with reduced number of iteration $r$ (line 8). Finally, the most efficient configuration is chosen (lines 12 and 13). 

\begin{algorithm}
\begin{algorithmic}[1]
\STATE{$E$ $\gets$ $\emptyset$, $V$ $\gets$ $\emptyset$}
\STATE{$L_{d}$ $\gets$ Algorithm \ref{alg:loop}}
\STATE{$L_{p}$ $\gets$ Fourier-Motzkin($L_{d}$)}
\STATE{$P_{t}$ $\gets$ check parallelism type($L_{d}$)}
\STATE{$L_{t}$ $\gets$ get loop transformations}
\FOR{$l_{t}$ $\textbf{in}$ $L_{t}$}
\STATE{$L$ = parallelise\_jvm($l_{t}$)} //Algorithm \ref{alg:jvm_p}
\STATE{$e$ = run($L$, $r$)}
\STATE{$E$ $\gets$ $E$ $\cup$ $e$}
\STATE{$V$ $\gets$ $V$ $\cup$ $L$}
\ENDFOR
\STATE{$id_{min}$ $\gets$ $arg$ $min$ $E$}
\RETURN{$V[id_{min}]$}
\caption{The parallelisation algorithm}
\label{alg:par}
\end{algorithmic}
\end{algorithm}

\section{Results}

\begin{figure}[!htbp]
  \centering
  \includegraphics[scale=1.25]{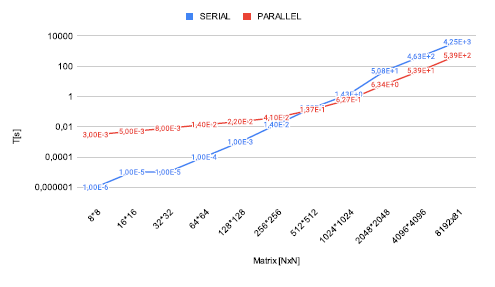}
    \caption{Logarithmic chart from serial and parallel time execution of matrix multiplication.}
  \label{fig:matrix_parallel_1}
\end{figure}

\begin{figure}[!htbp]
  \centering
  \includegraphics[scale=0.45]{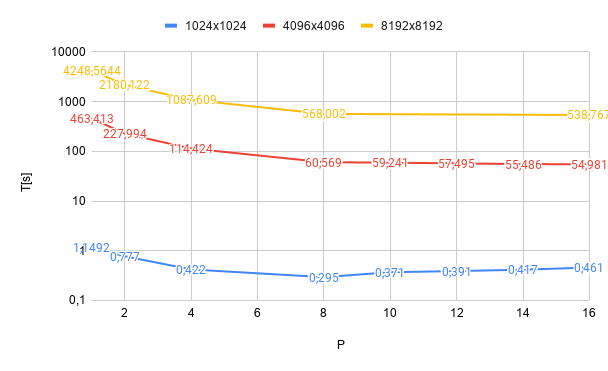}
    \caption{Logarithmic chart from matrix multiplication on different number of processor cores.}
  \label{fig:matrix_parallel_2}
\end{figure}

The presented framework was run on the following benchmarks: matrix multiplication, histogram computing, vanilla NBody problem and Fast Fourier Transform. The two parameters are efficiency and speedup are main indicators of parallelisation algorithm quality. The efficiency (E) is defined as: 
\begin{equation}
 E(N,P) = \frac{S(N,P)}{P} = \frac{T(N,1))}{P*T(N,P)}   
\end{equation}

and speedup (S):
\begin{equation}
S(N,P) = \frac{T(N,1)}{T(N,P)}    
\end{equation}

where: 
\begin{itemize}
    \item N - size of the problem, 
    \item P - number of cores, 
    \item T(N,P) - time execution for problem with size N and with P cores
\end{itemize}
In Table \ref{table:63} and Figures \ref{fig:matrix_parallel_1} and \ref{fig:matrix_parallel_2} the results for matrix multiplication are described. 

 \begin{table}[tpbh!]
\centering
\begin{tabular}{|l|l|l|l|l|l|l|l|l|l|} 
 \hline
 &
\multicolumn{3}{c|}{1024x1024} &
\multicolumn{3}{c|}{4096x4096} &
\multicolumn{3}{c|}{8192x8192} \\ [0.5ex] 
\hline\hline 
P & T[s] & E & S & T[s] & E & S & T[s] & E & S \\
\hline
1 & 1.15 & 1 & 1 & 463.41 & 1 & 1 & 4 248.56 & 1 & 1 \\
\hline
2 & 0.78 & 0.74 & 1.48 & 227.99 & 1.02 & 2.03 & 2 180.12 & 0.97 & 1.95 \\
\hline
4 & 0.42 & 0.68 & 2.72 & 114.42 & 1.01 & 4.05 & 1 087.61 & 0.98 & 3.91 \\
\hline
8 & 0.30 & 0.49 & 3.90 & 60.57 & 0.96 & 7.65 & 568.00 & 0.93 & 7.48 \\
\hline
10 & 0.37 & 0.31 & 3.10 & 59.24 & 0.78 & 7.82 &  &  &  \\
\hline
12 & 0.39 & 0.24 & 2.94 & 57.50 & 0.67 & 8.06 &  &  &  \\
\hline
14 & 0.42 & 0.20 & 2.76 & 55.49 & 0.60 & 8.35 &  &  &  \\
\hline
16 & 0.46 & 0.16 & 2.49 & 54.98 & 0.53 & 8.43 & 538.77 & 0.49 & 7.89 \\
\hline
\end{tabular}
\caption{Results of matrix multiplication with efficiency parameter E and acceleration S}
\label{table:63}
\end{table}

In Table \ref{table:63}, the results for different sizes of matrix are shown (1024x1024, 4096x4096 and 8192x8192). The efficiency is presented for serial and multicore version. The results are described for different numbers of cores. It can be observed that for smaller matrices (1024x1024) the peak performance is in the case of using eight cores. 

If the size is bigger (4096x4096 and 8192x8192) the best speedup is achieved while using all available cores - sixteen.
Figures \ref{fig:matrix_parallel_1} and \ref{fig:matrix_parallel_2} present these results using a logarithmic scale. Figure \ref{fig:matrix_parallel_1} shows comparison of execution times between serial and automatically generated parallel versions. It can be observed that around 256x256 size the generated code outperforms the serial one. 

\begin{figure}[!htbp]
  \centering
  \includegraphics[scale=0.8]{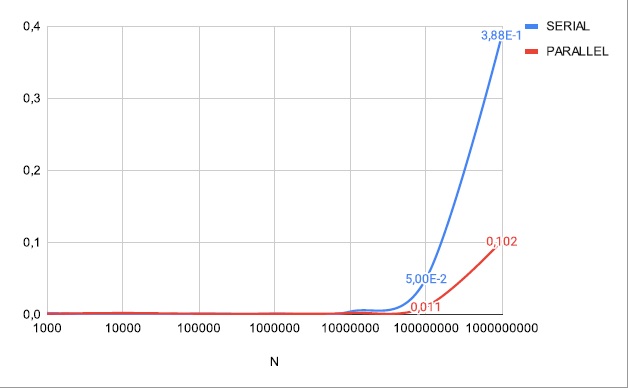}
    \caption{Scalability of histogram.}
  \label{fig:hist_1}
\end{figure}


\begin{figure}[!htbp]
  \centering
  \includegraphics[scale=1.55]{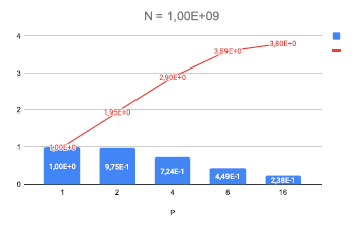}
    \caption{Histogram efficiency.}
  \label{fig:hist_2}
\end{figure}
\begin{figure}[!htbp]
  \centering
  \includegraphics[scale=1.3]{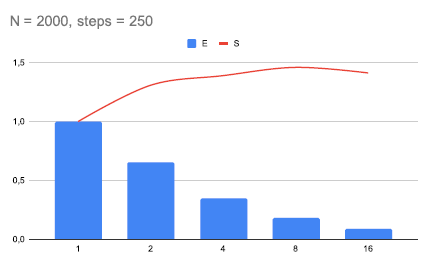}
    \caption{Nbody efficiency.}
  \label{fig:nbody}
\end{figure}

Figures \ref{fig:hist_1} and \ref{fig:hist_2} show histogram efficiency related to the size of input data and the number of cores. The histogram algorithm has data-driven parallelism. When the amount of data is about $10^7$ or higher then the acceleration can be noticed (Figure \ref{fig:hist_1}). Figure \ref{fig:nbody} presents Nbody efficiency. 

\begin{figure}[!htbp]
  \centering
  \includegraphics[scale=1.3]{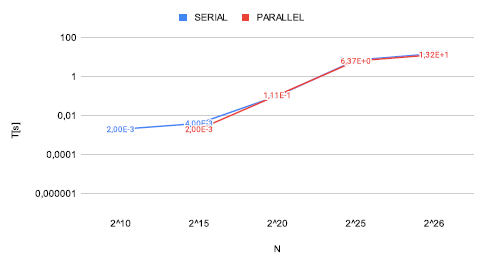}
    \caption{Logarithmic chart from FFT with different number of data points.}
  \label{fig:fft}
\end{figure}

It shows that maximum speedup achieved by automatically generated code was around 1.5 (for eight cores). Figure \ref{fig:fft} describes FFT scalability. 
The serial version is slightly more efficient than parallel. In all experiments the automatic parallel version is at the peak 15\% worse than its manually created counterpart.

All experiments were run on the processor Intel Core i9-9900K, 3.6GHz, RAM 16MB. Each experiment was repeated five times and average values were computed. The version of Java used in simulations was JDK 12.0. 



\section{Conclusions and future work}

Presented results show that described automatic translation algorithms can speedup various algorithms in java virtual machine. Moreover in many cases the generated parallel code can be as efficient as manually written code. Additionally, the depicted system can choose a proper accelerator and use appropriate strategy by using machine learning approaches. Future work will concentrate on further improvements in automatic parallelisation and testing JVM parallelisation modules on more languages like Scala, JRuby etc. 
New improvements will also concern machine learning techniques for execution parameter prediction and partial parallel code generation. Further work will also concentrate on testing more complex testbench algorithms for parallelisation.





%


\end{document}